\documentclass{PoS}

\title{Multiwavelength observations of blazars}

\ShortTitle{Multiwavelength observations of blazars}

\author{\speaker{Elena Pian}\\
        INAF-IASF Bologna, Via P. Gobetti 101, 40129 Bologna, Italy,\\
       and Scuola Normale Superiore di Pisa, Piazza dei Cavalieri 7, 56126 Pisa, Italy\\
        E-mail: \email{elena.pian@sns.it}}

\abstract{The {\it INTEGRAL} mission has played a major role in blazar science, thanks to  its sensitive coverage of a spectral region (3-100 keV) that is critical for this type of sources, to its flexibility of scheduling and to the large field of view of its cameras.  A number of flat-spectrum radio quasars (up to $z \sim 3$) and BL Lac objects were observed by {\it INTEGRAL} together with  facilities at all wavelengths.  These results have advanced our knowledge of blazars  from a physical and cosmological point of view. 
This paper reviews  some of these outcomes, with particular reference to the {\it INTEGRAL} program  for blazars in outburst as targets of opportunity, with a perspective into a  future of  multi-messenger astronomy.}

\FullConference{11th INTEGRAL Conference Gamma-Ray Astrophysics in Multi-Wavelength Perspective,\\
		10-14 October 2016\\
		Amsterdam, The Netherlands}

\begin{document}

\newcommand{\itbf}[1]{\textbf{\textit{#1}}}
\def\ltsima{$\; \buildrel < \over \sim \;$}
\def\simlt{\lower.5ex\hbox{\ltsima}} 
\def\gtsima{$\; \buildrel > \over \sim \;$}
\def\simgt{\lower.5ex\hbox{\gtsima}} 

\section{Introduction}

Blazars are the most luminous persistent (as opposed to gamma-ray bursts) gamma-ray emitters in the Universe.   Among extragalactic sources, they are the brightest   in the MeV-GeV range and the only detected ones at TeV energies (with the exception of a few very nearby radiogalaxies and  starbursts).  They represent more than 50\% of the significantly-detected sources in the 3rd Fermi-LAT catalog (Ackermann et al. 2015), and virtually the majority of the unidentified LAT targets.

The  typical "camel's back" (Falcke et al. 2004) shape of blazar spectral energy distribution is attributed to synchrotron radiation (responsible for emission at the lower frequencies) in a relativistic jet pointing at a small angle ($\simlt 5-10$ deg) with respect to the observer and inverse Compton scattering (at higher frequencies, up to GeV-TeV) of relativistic particles off synchrotron radiation or photon fields external to the jet  (accretion disk, broad emission line region, dusty torus).    The mutual relevance of external photon fields and  jet photons in  the inverse Compton scattering process determines the frequencies of the synchrotron and inverse Compton emission components peaks.  In flat-spectrum radio quasars, that have broad emission lines, the particles cool more effectively  than in BL Lac objects and the characteristic frequencies are accordingly lower  (Ghisellini et al. 1998).   

A range of continuous properties are observed between the most efficient coolers, flat-spectrum radio quasars with synchrotron peaks located at far-infrared wavelengths, and "extreme-synchrotron" BL Lac objects,  where the synchrotron peak is located at soft X-rays and occasionally moves, during outbursts, to hard X-rays (Pian et al. 1998).   These and the sources with intermediate properties represent a "blazar sequence" whereby total luminosities increase with decreasing characteristic frequencies (Fossati et al. 1998).  Notable exceptions to this picture are  a number of high-redshift blazars with broad luminous emission lines and high-frequency-peaked emission components (Arsioli et al. 2015).  

The phenomenology of the "blazar sequence" is naturally explained within a  scenario where the relativistic jet has a predominantly leptonic composition, which found confirmation in many studies of spectral energy distributions and multi-wavelength variability of blazars (e.g. Ackermann et al. 2012; Ghisellini et al. 2013; Marscher 2014; D'Ammando et al. 2015; Hayashida et al. 2015; Baring et al. 2017).  However, an alternative "hadronic" scenario cannot be discounted, and occasionally reproduces the observed spectral energy distribution in a more consistent way than the leptonic scenario, although it requires large powers in relativistic protons (B\"ottcher et al. 2013;  Diltz et al. 2015; Ackermann et al. 2016; Paliya et al. 2016).

\section{\itbf{INTEGRAL}  highlights in blazar research}

{\it INTEGRAL} detected a large number of blazars with the IBIS/ISGRI  instrument both in surveys  (Beckmann et al. 2009; Bassani et al. 2014; Krivonos et al. 2015; Malizia et al. 2016) and in targeted observations during flaring states, when the measurement of the hard X-ray spectrum (15-100 keV)  and variability  contributes, sometimes in a critical way, to the interpretation of the multi-wavelength observations (Courvoisier et al. 2003; T\"urler et al. 2006;  Foschini et al. 2006; Lichti et al. 2008; Vercellone et al. 2009; Collmar et al. 2010; Bottacini et al. 2010a).    Here I will concentrate on the results obtained with the program on blazars in outburst that  I have led since mission launch, and few more  specific results  led by collaborators.
 
In flat-spectrum radio quasars  and "intermediate BL Lac objects", the hard X-ray emission has a flat spectrum and is generally located below the spectral maximum of the inverse Compton component  (see Figure 1 in Falomo et al. 2014 and Fig. A1 in Ghisellini et al. 2011); therefore, it traces the behavior of the relativistic particles that produce the radio and infrared part of the synchrotron spectrum and helps locating the frequency of the inverse Compton maximum.  This was clearly measured in one of the first blazar sources observed by {\it INTEGRAL} IBIS, the high-redshift ($z = 2.172$) flat-spectrum radio quasar S5~0836+71, that was  detected in a high state  and with a very flat spectrum (photon index $\Gamma = 1.3$) during an {\it INTEGRAL} pointing of the 
intermediate BL  Lac S5~0716+71 (Pian et al. 2005).   This observation revealed one of the best advantages of {\it INTEGRAL} 
for investigation of the high energy sky, namely the large field of view of its cameras, that allow serendipitous detections of several  interesting targets (in that same pointing also two Seyfert galaxies were observed).

Among the brightest  blazars of the 3rd {\it Fermi}-LAT AGN catalog  (Ackermann et al. 2015), the flat-spectrum radio quasars  3C~279, 3C~454.3, 
PKS~1502+106,  and PKS~1510--089 were {\it INTEGRAL} targets in various occasions.  Our collaboration observed 
3C~454.3  ($z = 0.859$) in May 2005 at many frequencies other than hard X-rays, and found a rather different multi-wavelength spectral shape with respect to an earlier epoch, although the total energy had not varied significantly (Pian et al. 2006).  This made this source (which is also a strong {\it AGILE} source and TeV candidate, Vercellone et al. 2008) an excellent testbed for the "economic" jet model, whereby variations in individual bands can be conspicuous   (larger than a factor of 10), with the total energy  injected in the emitting region remaining however approximately constant (Katarzy\'nski \& Ghisellini 2007; Ghisellini et al. 2007).

PKS~1502+106 ($z = 1.839$) was observed by {\it INTEGRAL} in 2008 during a bright flare observed by {\it Fermi}-LAT.  IBIS did not detect it, although it set a significant constraint on the behavior of the inverse Compton spectrum  (Pian et al. 2011).  
Incidentally, one of the best studied Seyfert 1 galaxies, Mkn~841, is $\sim$8 arcmin away from PKS~1502+106 and detected by IBIS, so that we could compare its hard X-ray spectrum with prior observations and with a "classical" model including a soft X-ray excess, a Comptonized spectrum of a hot corona with  high energy cutoff and a reflected component (Bianchi et al. 2001).

The flat-spectrum radio quasar PKS~1510--089 ($z = 0.36$)  is well studied at all frequencies, including TeV (HESS Collaboration 2013),  and was repeatedly observed by   {\it INTEGRAL}.  We used this circumstance to assemble  data for this source over a rather long baseline (20 years) and to reconstruct its multi-wavelength history.   The spectral energy distribution was consistently compared with a model to trace the time-dependent behavior of the physical parameters and we searched the long data series for possible (quasi-)periodicities which were found however to be not significant  (Castignani et al. 2017).  

IBIS/ISGRI observed  3C~279 ($z = 0.539$),  the best known blazar, for 50 ks  during its June 2015  high state, as part of the {\it INTEGRAL} survey of the Coma region (Bottacini et al. 2016).  The broad-band spectral energy distribution  is consistent both with a  purely leptonic and a lepto-hadronic model, although the latter 
requires an extreme total jet power close to the Eddington luminosity of a black hole of mass similar to the one thought to lie at the center of 3C~279 ($\sim 2.5 \times 10^{8}$ M$_\odot$).   This high-state of 3C~279 adds to the rather complex multi-wavelength time behavior known from previous campaigns centered on {\it INTEGRAL} observations (Collmar et al. 2010).

Examples of the good synergy between {\it INTEGRAL} and  {\it XMM-Newton} are the high-redshift blazars PKS~0537-286 ($z = 3.1$) and PKS~2149--306   ($z = 2.345$)  whose  flat hard X-ray spectra made them  optimal targets for  {\it INTEGRAL}.  PKS~0537--286 was observed twice, in 2006 and 2008, and the IBIS spectrum constrained the modelling of the spectral energy distribution (Bottacini  et al. 2010b).   
PKS~2149--306 was observed by {\it INTEGRAL} IBIS and {\it Swift} BAT at  hard X-rays.  The comparison, together with {\it XMM-Newton} data, puts in evidence  two different states, that can be reproduced solely by a 
change of the bulk Lorentz factor of the jet  (Bianchin et al. 2009).

Particular interest is associated with the BL Lac object Mkn~421 ($z = 0.031$) both for its X-ray brightness and for the similarity of its overall spectral shape with Mkn~501, the prototypical "extreme synchrotron" blazar.  Hard X-ray observations of Mkn~421  in flaring state  have the potential to detect a shift of the synchrotron peak frequency of about two orders of magnitude, analogous to that detected in Mkn~501 (Pian et al. 1998), and establish this source as an equally powerful accelerator.    Our IBIS observation in April 2013, triggered by a high X-ray state seen by {\it Swift}-XRT and a simultaneous TeV flare seen by VERITAS, showed the source in a somewhat dimmer hard X-ray state than seen in 2006 (Lichti et al. 2008) and with steeper X-ray spectrum (Pian et al. 2014).   The synchrotron peak energy is limited to $\sim$1 keV, although {\it Swift} and {\it NuSTAR} observations in Jan-Jun 2013, bracketing the interval of our IBIS campaign, detected it to vary between 1 and 10 keV (Kapanadze et al. 2016).   The  simultaneous {\it  INTEGRAL} IBIS and JEM-X light curves in April 2013 indicate complex variability with direct correlation at the various frequencies, and a possible delay that increases monotonically with energy difference and reaches  about 1 hour between the fluxes at the lowest  (3-5 keV) and  the highest (40-100 keV) frequencies  (Pian et al. 2014).

\section{Conclusion and future prospects}

The simultaneous use of {\it INTEGRAL} and other  space-  and ground-based multi-wavelength facilities has led to substantial progress in mapping the behavior of jets both for BL Lac objects and for emission line radio quasars, owing to its  sensitivity in a critical interval of the blazar spectrum.    This is due both  to its  flexibility  of scheduling and repointing at blazar targets of opportunity  following alerts from other missions (see Figure 1), and to the large field-of-view  of its  cameras that  allows serendipitous detection of blazars in active state  (see Bottacini et al. 2010a, 2016).

{\it INTEGRAL}'s blazar  legacy  is bestowed on future space missions and ground-based telescope networks that will  refine the level of coordination, coverage, monitoring duration and sampling rate.   Optimization of blazar campaigns will lead to 
understanding the details of blazar jets structure and physics.  In particular, it is becoming increasingly clear that the geometry of the  emitting region is complex, so that a homogeneous region approximation may not be valid for an accurate description of multi-wavelength variability.  Moreover, different jet composition,  leptonic vs lepto-hadronic, may play a critical role  (B\"ottcher 2010).

Blazar investigation will benefit from the addition of the important  multi-messenger  aspect  represented by ultra-high energy neutrinos.  IceCube has so far detected a few extremely energetic events above 100 TeV,  one of which  was recently proposed to be associated  with a gamma-ray flare  detected by {\it Fermi}-LAT in blazar PKS B1424-418 at $z = 1.522$ (Kadler et al. 2016).
While neutrino detection  may favour  a hadronic scenario for blazars, a structured  jet as envisaged  in a   spine-and-sheath geometry may   also be viable (Tavecchio \& Ghisellini 2015).

\begin{figure}
\includegraphics[width=1.0\textwidth]{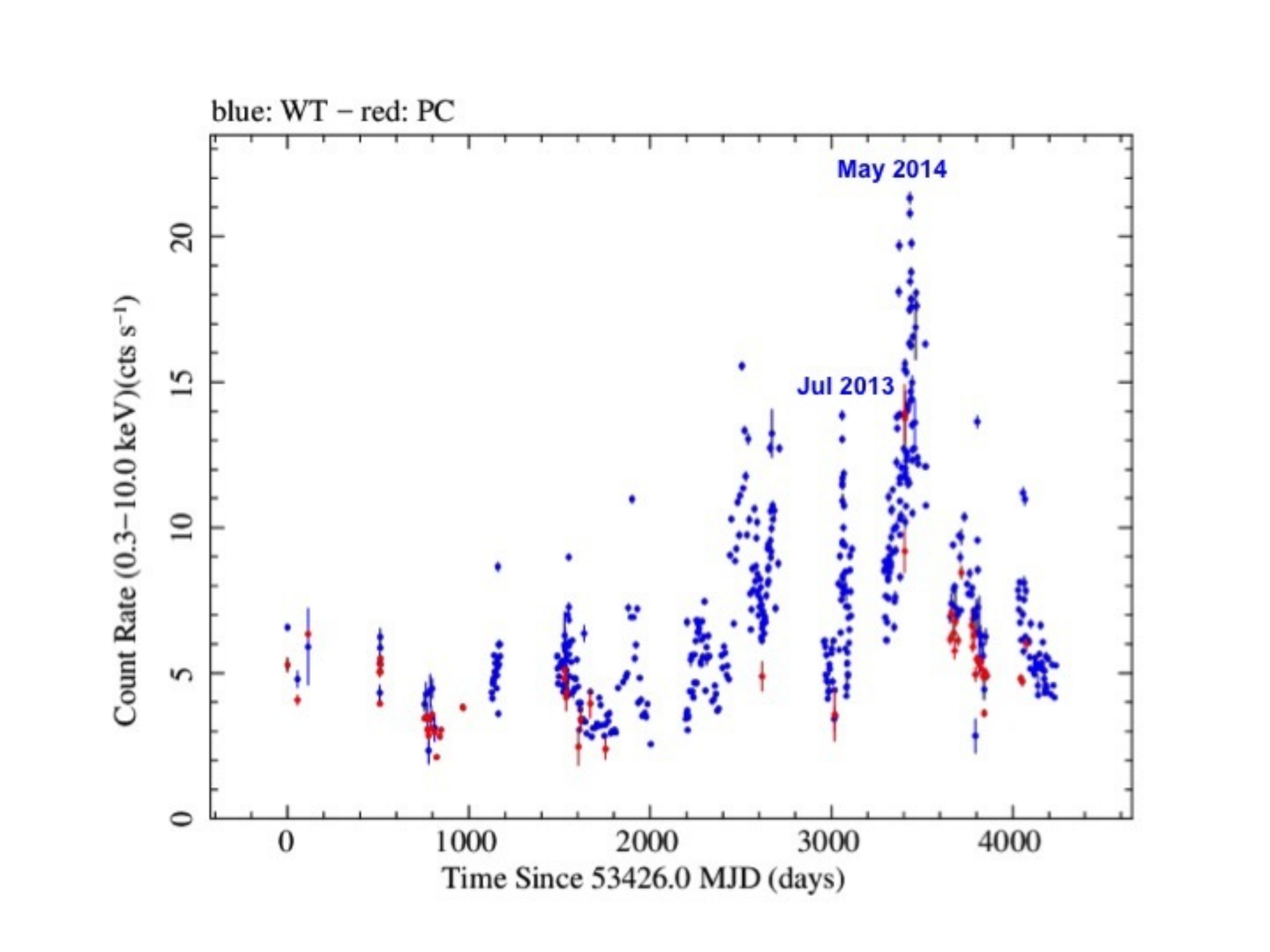}
\caption{{\it Swift}/XRT light curve (from http://www.swift.psu.edu/monitoring/) of the BL Lac object Mkn~501 ($z = 0.034$),  between 2005 February and 2015 July.  In the abscissa, MJD = 0 corresponds to  2005 Feb 25.0 UT.  The two flares of July 2013 and  May 2014  are indicated: at these epochs, {\it INTEGRAL} observations were accomplished (E. Pian et al., in preparation).}
\label{fig1}
\end{figure}

\section{Acknowledgements}

I am indebted to all collaborators who have helped with the success of the blazar {\it INTEGRAL} program with  inputs at so many different levels:
P. Barr, A. Bazzano, V. Beckmann, T.M. Belloni, S. Bianchi, V. Bianchin, M. B\"ottcher, R. Boissay, E. Bottacini, G. Castignani, S. Ciprini, W. Collmar, T. Courvoisier, F. D'Ammando, G. Di Cocco, D. Eckert, C. Ferrigno, M.T. Fiocchi, L. Foschini,  L. Fuhrmann,  N. Gehrels, G. Ghisellini, P. Giommi, R. Hudec, D. Impiombato, E. Lindfors, G. Malaguti, L. Maraschi, A. Marcowith, P. Michelson, K. Nilsson, G. Palumbo, M. Pasanen, M. Persic, T. Pursimo,  C.M. Raiteri, P. Romano, T. Savolainen, M. Sikora, A. Sillanp\"a\"a, S. Soldi, A. Stamerra, G. Tagliaferri, L. Takalo, F. Tavecchio, D. Thompson, M. Tornikoski, G. Tosti, A. Treves, M. T\"urler, P. Ubertini, E. Valtaoja, S. Vercellone, M. Villata,  R. Walter, and A. Wolter.  I  am  grateful also to E. Kuulkers, P. Kretschmar, G. Belanger and the staff at {\it INTEGRAL} Science Operations Center and Science Data Center for their support of  the  program.  Finally, I thank   the  organizers for a memorable conference,  nice  tour of Amsterdam and excellent Chinese banquet.

\end{document}